\documentclass[a4paper,11pt]{article}
\usepackage{pos}
\usepackage{todonotes,siunitx,xfrac}
\usepackage{multirow,threeparttable,makecell}
\usepackage[normalem]{ulem}

\usepackage{graphicx,booktabs}
\usepackage{tikz,atbegshi}
\usepackage{subfigure}
\usetikzlibrary{calc}
\usepackage{textcomp}
\usepackage{multirow}
\usepackage{siunitx}
\usepackage{makecell}
\usepackage{caption} 
\usepackage{wrapfig}
\newcommand{\eg}{\textit{e.g., }}

\usepackage{xcolor}%

\newcommand{\pbar}{$\bar{p}$}
\newcommand{\eV}{\si{\electronvolt}}
\newcommand{\keV}{\si{\kilo\electronvolt}}
\DeclareSIUnit\bar{bar}
\usepackage{orcidlink}

\title{Towards Precision Spectroscopy of Antiprotonic Atoms for Probing Strong-field QED}

\author[1]{G.~Baptista\orcidlink{0000-0002-0375-3433}}
\author[1,10]{S.~Rathi\orcidlink{0000-0002-0768-5546}}
\author[1]{M.~Roosa\orcidlink{0000-0002-3449-4665}}
\author[1]{Q.~Senetaire\orcidlink{0009-0008-4437-2099}}
\author[1]{J.~Sommerfeldt\orcidlink{0000-0002-3471-7494}}

\author[2]{T.~Azuma\orcidlink{0000-0002-6416-1212}}

\author[3]{D.~Becker\orcidlink{0000-0001-6942-8636}}
\author[4]{F.~Butin\orcidlink{0000-0003-3068-3343}}
\author[10]{O.~Eizenberg\orcidlink{0009-0000-0489-1582}}
\author[3]{J.~Fowler\orcidlink{0000-0002-8079-0895}}
\author[5]{H.~Fujioka\orcidlink{0000-0001-8653-3761}}
\author[4]{D.~Gamba\orcidlink{0000-0002-1985-1847}}
\author[6]{N.~Garroum}
\author[7]{M.~Guerra\orcidlink{0000-0001-6286-4048}}
\author[8]{T.~Hashimoto\orcidlink{0000-0001-6973-4895}}
\author[9]{T.~Higuchi\orcidlink{0000-0003-3281-4669}}
\author[1]{P.~Indelicato\orcidlink{0000-0003-4668-8958}}
\author[7]{J.~Machado\orcidlink{0000-0002-0383-4882}}
\author[3]{K.~Morgan\orcidlink{0000-0002-6597-1030}}
\author[1]{F.~Nez\orcidlink{0000-0002-3478-7521}}
\author[3]{J.~Nobles\orcidlink{0000-0002-7538-0031}}
\author[10]{B.~Ohayon\orcidlink{0000-0003-0045-5534}}
\author[11]{S.~Okada\orcidlink{0000-0002-4465-9961}}

\author[3]{D.~Schmidt\orcidlink{0000-0002-5764-0194}}
\author[3]{D.~Swetz\orcidlink{0000-0002-8192-2175}}

\author[3]{J.~Ullom\orcidlink{0000-0003-2486-4025}}
\author[1]{P.~Yzombard\orcidlink{0000-0002-0864-181X}}
\author[6]{M.~Zito}

\author[1]{N.~Paul\orcidlink{0000-0003-4469-780X}}

\affiliation[1]{Laboratoire Kastler Brossel, Sorbonne Université, CNRS, ENS-PSL Research University, Collège~de~France, Case 74; 4, place Jussieu, F-75005 Paris, France}
\affiliation[2]{RIKEN Atomic, Molecular and Optical Physics Laboratory, RIKEN, Wako, Saitama 351-0198, Japan}

\affiliation[3]{National Institute of Standards and Technology, Boulder, Colorado 80305, USA}
\affiliation[4]{CERN, Meyrin, Switzerland}
\affiliation[5]{Institute of Science Tokyo, Tokyo 152-8551, Japan}
\affiliation[6]{Laboratoire de physique nucléaire et des hautes énergies/Institut national de physique nucléaire et de physique des particules, 4, place Jussieu, 75005, Paris, France}
\affiliation[7]{Laboratory of Instrumentation, Biomedical Engineering and Radiation Physics (LIBPhys-UNL), Department of Physics, NOVA School of Science and Technology, NOVA University Lisbon, 2829-516 Caparica, Portugal}
\affiliation[8]{RIKEN Nishina Center, RIKEN, Wako 351-0198, Japan}
\affiliation[9]{Institute for Integrated Radiation and Nuclear Science, Kyoto University, Osaka 590-0494, Japan  }
\affiliation[10]{Physics Department, Technion—Israel Institute of Technology, Haifa 3200003, Israel}
\affiliation[11]{Department of Mathematical and Physical Sciences, Chubu University, Kasugai, Aichi 487-8501, Japan}

\emailAdd{npaul@lkb.upmc.fr}

\abstract{
\textbf{Abstract:} PAX (antiProtonic Atom X-ray spectroscopy) is a new experiment with the aim to test strong-field quantum electrodynamics (QED) effects by performing high-precision x-ray spectroscopy of antiprotonic atoms. 
By utilizing advanced microcalorimeter detection techniques and a low-energy antiproton beam provided by the ELENA ring at CERN, gaseous targets will be used for the creation of antiprotonic atoms, and the measurement of transitions between circular Rydberg states will be conducted with up to two orders of magnitude improved accuracy over previous studies using high-purity germanium detectors. 
Our approach eliminates the longstanding issue of nuclear uncertainties that have hindered prior studies using highly charged ions, thus enabling direct and purely QED-focused measurements. 
By precisely probing atomic systems with electric fields up to two orders of magnitude above the Schwinger limit, PAX will test vacuum polarization and second-order QED corrections, opening new frontiers in fundamental physics and uncovering potential pathways to physics beyond the Standard Model.

}

\begin{document}
\maketitle
\section{Introduction}

As the best-understood quantum field theory, quantum electrodynamics (QED) offers a promising avenue for exploring the details of the Standard Model and seeking hints of new physics \cite{2018-Safranova}. The significant progress made in precision measurements of atomic systems means that bound-state quantum electrodynamics (BSQED) can be tested, wherein atoms serve as a unique laboratory for probing the details of the quantum vacuum. 

For simple weak-field  low-$Z$ systems, such as the hydrogen atom, extremely high-precision BSQED calculations can be performed perturbatively in the Coulomb field ($\alpha Z$), allowing for a comparison with laser-spectroscopy experiments at the level of 12 to 15 significant digits~\cite{Matveev2013,Fleurbaey2018} and tests up to third order interactions with the vacuum. However, when moving to higher-$Z$, a fundamentally different approach must be used, as the interactions with the Coulomb field can no longer be treated perturbatively, and so-called all-order BSQED calculations must be performed in this strong field regime. These all-order calculations have been much less stringently tested due to a multitude of experimental and theoretical challenges \cite{Indelicato2019}. 
This is unfortunate, as the field strengths in high-$Z$ systems approach the Schwinger limit - the threshold at which the electromagnetic field strength exceeds $\sim1.32\times10^{18}~\si{\volt\per\meter}$, where in principle real electron-positron pair production becomes possible - which enhances both relative contributions from BSQED effects in the bound-state atomic structure and the potential to observe new effects in the electromagnetic regime. 

Precise strong-field BSQED tests can only be performed in systems that are simultaneously simple enough to be accessible to theory and extreme enough to achieve the field strengths of interest. 
As such, many recent studies have focused on high-precision measurements of x rays coming from inner-shell atomic transitions in few-body, highly charged ions (HCIs) across a broad range of $Z$ \cite{Machado2023, Duval2024,Trotsenko2010, Gumberidze2007, Loetzsch2024}; 
see also \cite{Indelicato2019} and references therein.
However, the stripping of high-$Z$ atoms is a challenging task, necessitating their spectroscopy to be performed in plasmas with electron cyclotron ion sources, or in storage rings, both of which produce limiting Doppler shifts. 
For instance, a measurement of the Lamb shift in hydrogen-like (H-like) uranium with a germanium detector was limited to an experimental precision of 10$^{-3}$ \cite{Gumberidze2005}, which is insufficient for second-order QED tests. 
A more fundamental roadblock to producing strong-field BSQED benchmarks through spectroscopy of HCIs is the poorly constrained nuclear properties of these heavy systems, which introduce significant uncertainties in transition energies~\cite{Indelicato2019}, limiting the sensitivity of the measurements to only first-order QED. 

\subsection{The PAX Methodology}

\begin{figure}[htb!]
    \centering
    \includegraphics[width=0.6\linewidth]{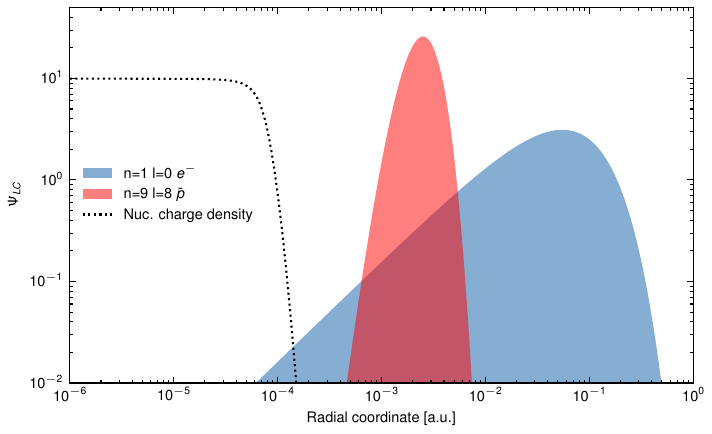}
    \caption{The radial part of selected electronic (blue) and antiprotonic (red) wavefunctions (large component) in $^{40}$Ar are shown. The dashed line represents the nuclear charge distribution, with parameters obtained from previous electron scattering experiments~\cite{DEVRIES1987495}, and has been normalized for visualization purposes.}
    \label{fig:Ryd_wf}
\end{figure}

PAX's approach to strong-field BSQED studies addresses the above challenges by using a new paradigm of studying Rydberg transitions in antiprotonic (\pbar) atoms \cite{Paul2021}. In these systems, composed of a nucleus and an antiproton, the heavy mass of the antiproton ($\approx$1836 times the mass of the electron) leads to Bohr radii that are significantly smaller than the electronic equivalents, resulting in much stronger field intensities. 
Even for high-\textit{n} antiprotonic Rydberg states, the Coulomb field strengths can be orders of magnitude higher than for a 1\textit{s} electronic state in their H-like counterpart (\eg three orders of magnitude for $n=6$ in \pbar Ne and two for $n=11$ in \pbar Pb). This opens up the possibility of studying transitions between circular Rydberg states, where the wavefunctions are nodeless and much more spatially confined, providing a minimal overlap with the nuclear core and effectively eliminating the uncertainty contributions from the nuclear structure. For a clearer picture, Figure~\ref{fig:Ryd_wf}  illustrates the comparison between an antiprotonic circular Rydberg state (\textit{n}$=9$, \textit{l}$=8$) and a 1s electron in H-like Ar.
Further, Table~\ref{table:QEDcomp} highlights the magnified QED effects in \pbar\ systems by comparing a Rydberg transition in \pbar Xe with the Lyman-$\alpha_1$ transition in H-like U, falling within the same energy range. The first- and second-order QED effects in \pbar Xe are 1.5 and 3 times larger, respectively, than those in Lyman-$\alpha_1$ of H-like U. In contrast, finite nuclear size (FNS) effects in \pbar Xe are 100 times smaller, significantly below the second-order QED, whereas in H-like~U, FNS contributions are of the same order of magnitude as the first-order QED contributions.


\begin{table}[h!]
\centering
\begin{threeparttable}
\caption{Comparison of QED and nuclear contribution in \pbar Xe and H-like U (U$^{91+}$)} \label{table:QEDcomp}
\begin{tabular}{c|ccc|c|c|c|c|c}
\toprule

& & & &\thead{Energy}  & \thead{1$\textsuperscript{st}$  order QED} & 
\thead{2$\textsuperscript{nd}$  order QED} & \thead{FNS} & \thead{FNS/QED} \\
\multirow{-2}{*}{\thead{Element}}&\multicolumn{3}{c|}{\multirow{-2}{*}{\thead{Transition}}}&\thead{[$\eV$]}&\thead{ [$\eV$]}&\thead{ [$\eV$]} &\thead{ [$\eV$]} &\thead{ [\%]} \\\midrule

\multirow{1}{*}{\pbar Xe} & 12o$_{\sfrac{21}{2}}$&$\rightarrow$&11n$_{\sfrac{21}{2}}$  &95618.8&398.9&3.9&0.8 & 0.198 \\
\midrule
\multirow{1}{*}{{U$^{91+}$}}&2p$_{\sfrac{3}{2}}$&$\rightarrow$&1s$_{\sfrac{1}{2}}$& 102175.1& $-257.2$&1.2&$-198.5$ & 77.173\\
\bottomrule
\end{tabular}

\end{threeparttable}

\end{table}

Exploiting these unique characteristics of antiprotonic atoms, PAX aims to measure transitions at sufficiently high principal quantum numbers where antiproton annihilation with a nucleon is unlikely to occur until a later stage. At these quantum states, strong-force shifts and broadenings are negligible, while FNS corrections are smaller than second-order QED effects, allowing the latter to become experimentally accessible for the first time in these systems.

PAX plans to study elements of a broad range of $Z$ in gaseous targets. 
These conditions minimize the probability of electron refilling, reducing uncertainties associated with poorly understood atomic structures and avoiding the resulting increase in spectral complexity~\cite{Gotta2008}. An example of targeted transitions for the PAX physics program is shown in Table~\ref{table:trans}.


Although antiprotonic atoms have been extensively studied in the past, notably at the now decommissioned LEAR facility at CERN, these pioneering experiments were mostly focused on accurate measurements of antiprotonic x rays originating from close-to-nucleus transitions (lower $n$), where strong-interaction effects are boosted~\cite{Kreissl1988,Kreissl1988_00,Daniel1987,WYCECH1993607}. Other efforts also studied transitions from higher-$n$ circular Rydberg states, such as for the measurement of the magnetic moment of the antiproton~\cite{Kreissl1988}; or even for QED purposes, but the employment of germanium detectors added a limitation of their in-beam detector resolution of $\frac{\Delta E}{E} = 0.04$ \cite{Gotta2008}. 

\begin{figure}[htb!]
    \centering
    \includegraphics[width=0.6\linewidth]{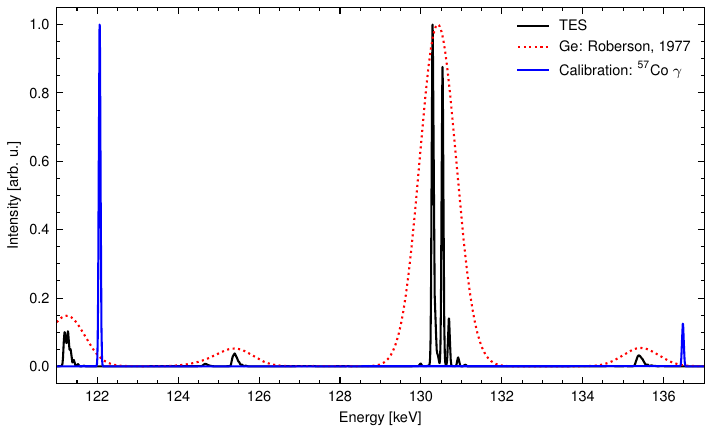}
    \caption{Theoretical spectrum of antiprotonic $^{40}$Zr x rays. The black and red lines represents the expected measurement when performed with a TES (50~$\eV$ resolution) and a Ge detector (1.03~$\keV$ resolution at 123~$\keV$~\cite{PhysRevC.16.1945}), respectively, of antiprotonic x rays. The better resolution of a TES allows for the identification of the fine structure splittings in the $(n=9,\ l=8)\rightarrow(n=8,\ l=7)$ transitions, as well as observing parallel transitions (from non-circular Rydberg states). No background was accounted for in this simulation. Hadronic shifts and broadenings, due to strong interaction, are not expected to be observed in the plotted transitions. The $122.060\, 65\, (12)\ \keV$ and $136.473\, 56\, (29)\ \keV$ gamma lines from $^{57}$Co \cite{Helmer2000} (in blue) are well-known candidates for calibrating this measurement, due to their sub-eV accuracy measurements and absence of natural broadening.}
    \label{fig:calibration}
\end{figure}

Experimentally, the key to attaining the physics goals of the PAX  program lies in replacing the germanium detectors used at LEAR ~\cite{Kreissl1988,Kreissl1988_00,Daniel1987,WYCECH1993607,Gotta2008,Simons1993} with a state-of-the-art microcalorimeter detector, a novel quantum sensor, as also used by sister-collaborations QUARTET~\cite{physics6010015,Unger2024} and HEATES~\cite{Okada2020,Okumura2021,Okumura2023,Okumura2024,Saito2025} for muonic atoms (composed of a nucleus and a bound negative muon). X-ray microcalorimeters are revolutionizing the field of x-ray and low-energy gamma ray spectroscopy~\cite{Ullom2015,Doriese2017}. These detectors, based on either superconducting (transition edge sensors - TESs) or magnetic transitions (magnetic microcalorimeters - MMCs)~\cite{Ullom2015, Pies2012, Fleischmann_2005} offer, for the first time, both high resolution ($\sim$10$^{-4}$ intrinsic resolution) and high detection efficiency ($\sim$0.4 quantum efficiency), making possible the precision spectroscopy of exotic species, where luminosity is limited. 
The efficacy of these detectors in muonic atom studies has already been established at JPARC~\cite{Okada2020,Okumura2021,Okumura2023}, thus demonstrating their compatibility with the harsh environments present at accelerator facilities. The choice to use a Transition Edge Sensor (TES) for PAX
will allow to perform measurements with maximal solid angle and an intrinsic resolution of about $50\ \eV$ (FWHM)  in the $50$--$250\ \keV$ range, with an expected $10^{-5}$--$10^{-6}$ accuracy, should the proper calibration scheme be employed so as to counteract the characteristic nonlinearities of these detectors. Figure~\ref{fig:calibration} exemplifies the significant gain in intrinsic resolution expected with the PAX detector, compared to a previous solid-target measurements with a Ge detector.  

\begin{table}
\centering
\begin{threeparttable}
\caption{Theoretical calculations for the antiprotonic transitions targeted for PAX's final QED tests. In these systems, the FNS contributions, while still comparable, are lower than the 2$^{\text{nd}}$ order effects, and the measurements shall not be limited by the expected sub-$\eV$ accuracy. More on how these calculations were conducted can be found in Section~\ref{sec:theory}.} \label{table:trans}
\begin{tabular}{c|lcl|c|c|c|c|c}
\toprule

& & & &\thead{Energy}  & \thead{1$\textsuperscript{st}$  order QED} & 
\thead{2$\textsuperscript{nd}$  order QED} & \thead{FNS}&\thead{FNS/QED} \\
\multirow{-2}{*}{\thead{Element}}&\multicolumn{3}{c|}{\multirow{-2}{*}{\thead{Transition}}}&\thead{[$\eV$]}&\thead{ [$\eV$]}&\thead{ [$\eV$]} &\thead{ [$\eV$]}&\thead{[\%]} \\\midrule

\multirow{1}{*}{$^{20}$Ne}& 6h$_{\sfrac{11}{2}}$&$\rightarrow$&5g$_{\sfrac{9}{2}}$  &29\ 173.9&106.4&1.0&0.07&0.065 \\
\midrule

\multirow{1}{*}{$^{40}$Ar} & 6h$_{\sfrac{11}{2}}$&$\rightarrow$&5g$_{\sfrac{9}{2}}$  &$97\ 002.1$&524.5&5.2&1.1&0.208 \\
\midrule
 \multirow{1}{*}{$^{132}$Xe}&10m$_{\sfrac{19}{2}}$&$\rightarrow$&9l$_{\sfrac{17}{2}}$&170\ 495.2&906.0&9.1&3.2&0.350\\ \midrule
\multirow{1}{*}{$^{184}$W}&12o$_{\sfrac{23}{2}}$&$\rightarrow$&11n$_{\sfrac{21}{2}}$&180\ 553.1&912.2&9.3&3.6&0.391\\
\bottomrule
\end{tabular}

\end{threeparttable}
\end{table}

\section{Theory} \label{sec:theory}

In order to utilize the full potential of PAX as a test of strong-field BSQED, accurate theoretical predictions for the transition energies that match the experimental precision are needed. These theoretical values will be obtained from state-of-the-art MCDF computations using the \textit{MCDFGME} code~\cite{Desclaux1975,Indelicato1990,Indelicato2005,Indelicato2013} that account for the interaction of the antiproton with the atomic nucleus. Moreover, we compute all QED effects that are expected to be relevant, which requires careful treatment of vacuum polarization (VP) effects as their importance is significantly enhanced in antiprotonic atoms compared to their electronic counterparts \cite{Paul2021}. We will compute the first- and second-order VP corrections of the energy levels, including corrections from the finite nuclear and antiprotonic size. Additionally, the first-order self-energy (SE) and mixed SE and VP corrections will be included in the theoretical predictions, since they are expected to exceed the desired accuracy of PAX. Methods for computing these QED contributions are either already available and included in the preliminary theoretical values of Table~\ref{table:QEDcomp} and \ref{table:trans} or are currently being adapted from electronic to antiprotonic atoms, see Reference ~\cite{Paul2021} for more details on the QED effects currently taken into account.

\section{Experimental Apparatus}\label{sec:experiment}

\begin{figure}[htb]
    \centering
    \begin{minipage}{0.55\textwidth}
        \centering
        \includegraphics[width=1\linewidth]{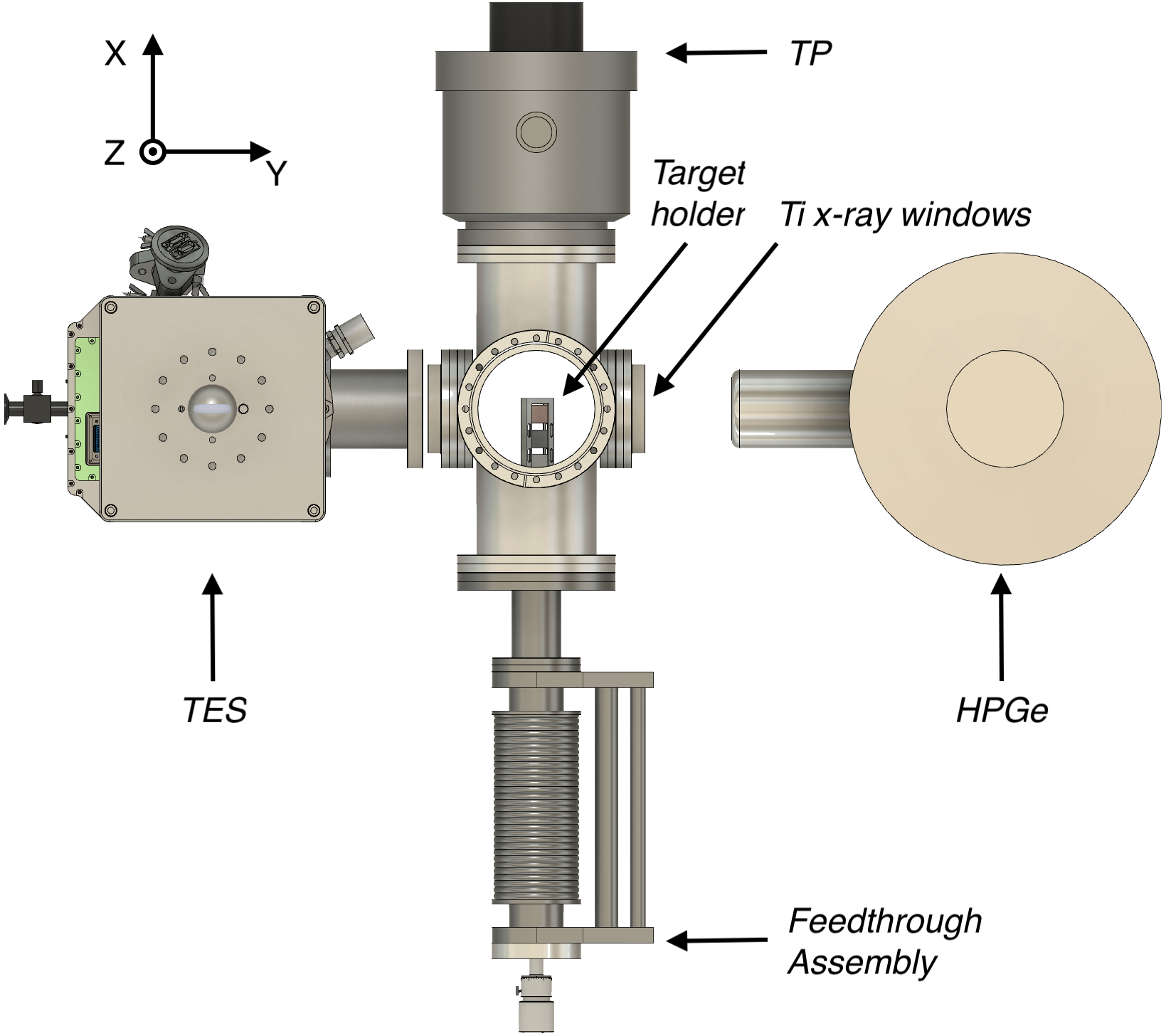}
    \end{minipage}
    \begin{minipage}{0.44\textwidth}
        \centering
        \includegraphics[width=1\linewidth]{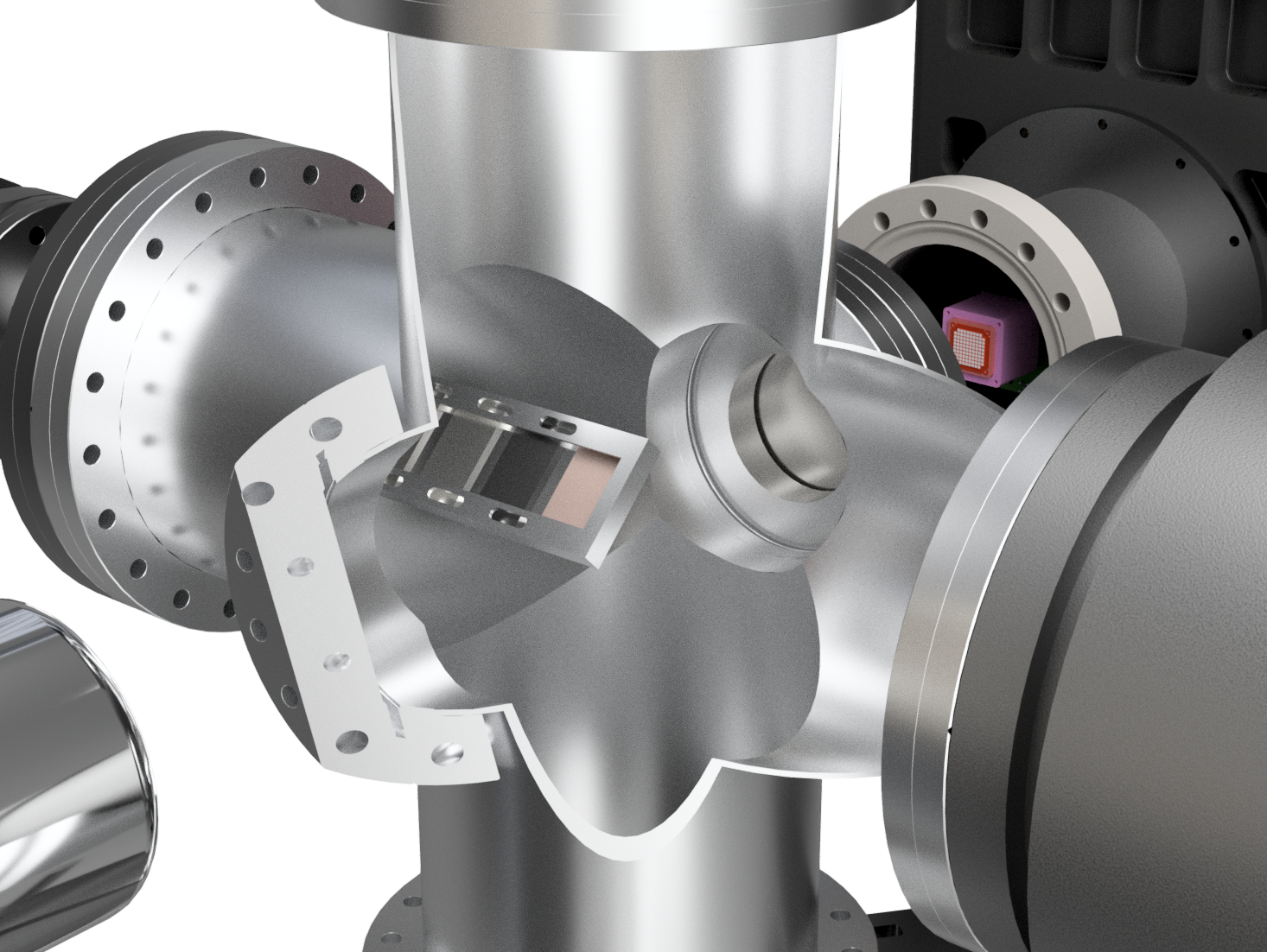}
    \end{minipage}
    \caption{Overview of the experimental setup for the PAX test beam for solid targets. (Left) The z-axis depicts the vertical. 100 keV antiprotons arrive along the z direction onto solid targets attached to a ladder. The target holder is connected to a motion feedthrought assembly made of a \SI{20}{\cm} stroke linear translator along with a rotary feedthrough, allowing a 45 degrees orientation of the target with respect to the z-x plane. Two turbomolecular pumps (TP) coupled in series attached at one lateral port of the chamber allow to achieve \SI{\approx 8E-10}{\milli\bar} vacuum. Two thin titanium windows installed on the lateral flanges allow the x rays from the target to be measured by the detectors. The top flange will be equipped with a diagnostic detector to monitor the antiproton beam parameters. The prototype TES detector will stand on a translation-cart with a \SI{.5}{\meter} stroke, enabling a tunable solid angle. A high-purity Germanium (HPGe) detector will measure x rays that pass through the targets, serving as a diagnostic. (Right) Alternative view of the experimental setup.}
    \label{fig:setup}
\end{figure}

The PAX experimental scheme consists of impinging 100 keV antiprotons from the Extra Low ENergy Antiproton (ELENA) ring at CERN onto a target cell, and detecting the emitted x rays from the antiprotonic atom cascade with a novel, large-area TES detector. 
While gas targets are needed for probing QED effects in the final physics program, the first test beam will use solid targets for benchmarking the detector's performance and to compare with previous measurements. 
As this will be the first-ever deployment of a microcalorimeter detector in conjunction with antimatter beams, a careful assessment and control of systematic effects are essential. The experimental setup for a solid-target configuration is shown in Figure~\ref{fig:setup}. A germanium detector will measure in parallel with the TES for monitoring. As evident in Figure~\ref{fig:setup}, the solid-target configuration will feature an ultra-high vacuum (UHV) target chamber and a translating target ladder. 
The gas target configuration will see the UHV chamber replaced by a mbar-range gas cell sealed behind an aluminized mylar window, as is common practice at ELENA~\cite{Nordlund2022,Latacz2023}.
The TES will be positioned orthogonal to the beam axis in both cases. Gamma-ray calibration sources deposited on thin windows will be placed between the interaction region and the TES to provide continuous calibration lines that uniformly illuminate the pixel array.

\begin{figure}[htb]
    \centering
    \includegraphics[height=0.18\textheight]
    {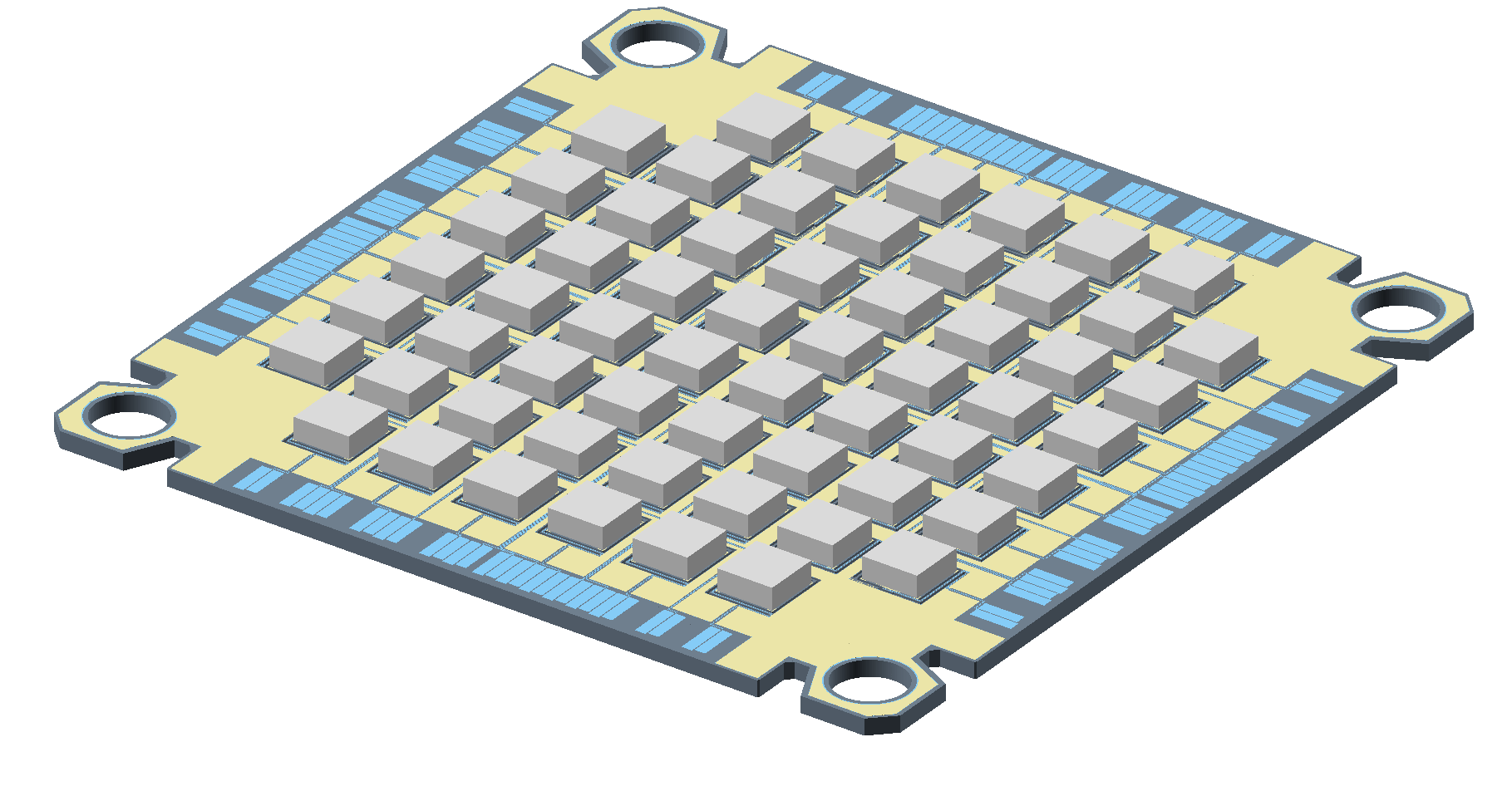}
    \hspace{.5cm}
    \includegraphics[height=0.18\textheight]
    {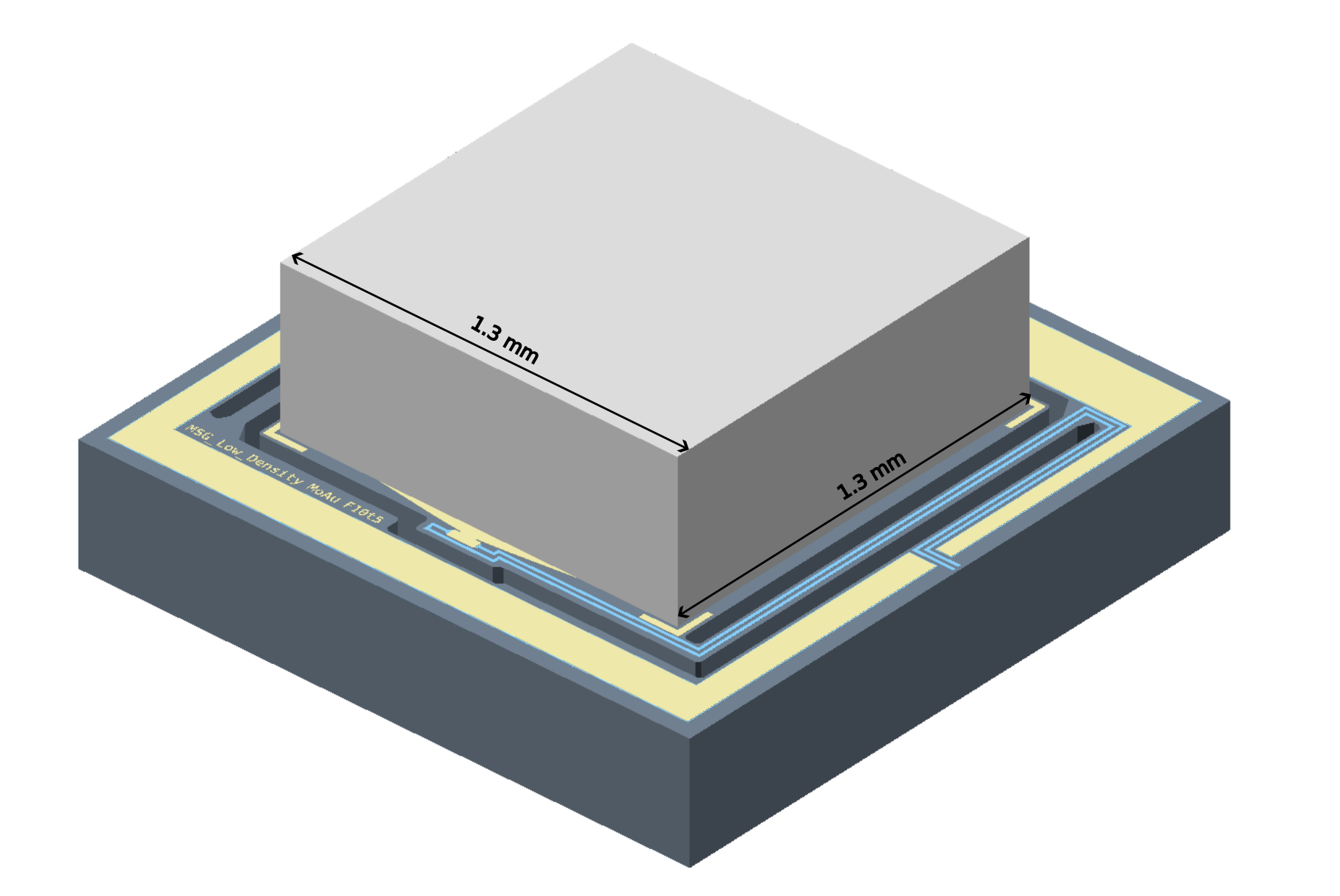}
    \caption{Renderings of the absorber array of one PAX microsnout (Left) and single-pixel (Right) are shown. The chip holding the absorber array is $2\ \si{\centi\meter}\times2\ \si{\centi\meter}$. The Sn absorber is drawn in light grey, and the Mo readout wires are in blue. The dark grey and yellow correspond to Si and Au, respectively. The final detector will combine four microsnouts for a total of 240 pixels.
    }
    \label{fig:tes-drawing}
\end{figure}
The major experimental development for PAX is a novel large-area TES x-ray microcalorimeter being developed with the National Institute of Standards and Technology (NIST) Quantum Sensing Division, USA. 
A prototype design is being developed for an in-beam test with solid targets scheduled for spring 2025, consisting of a tin-based x-ray absorber on a pure silicon substrate with an active area of approximately 9 cm$^2$.  
The first in-beam test will be used to optimize the absorber design in view of building a larger area TES array that will be deployed for PAX's final measurements. The final detector design will feature a group of four pixel arrays (called microsnouts), with 96 pixels each of $1.3\ \si{\milli\meter}\times1.3\ \si{\milli\meter}$, for a total active area of \SI{650}{\milli\meter^2}. The pixels will be based on tin (Sn) absorbers with thicknesses between 0.38 and \SI{0.5}{\milli\meter} optimized for best resolution at around 100 keV, but with the capability of detecting x rays in the 30--250~$\keV$ range. 
SQUIDs (Superconducting QUantum Interference Device) will be used for the sensor readout, with a microwave SQUID multiplexing scheme \cite{Mates2008,Kempf2014,Mates2017}. 
A dilution refrigerator would be too large, so the project is based on a $^{3}$He-backed adiabatic demagnetization refrigerator. 
A preliminary design of the active parts of the detector is shown in Figure~\ref{fig:tes-drawing}.

\section{Antiproton beam structure}

Achieving the lower bound accuracy goals of the PAX program ($\frac{\Delta E}{E}\approx 10^{-5}$)  will necessitate the reduction of the instantaneous antiproton rates from the current standard ELENA deliverable of one single \SI{100}{\nano\second} FWHM bunch of $10^{7}$ antiprotons every 2 minutes. 
After taking into consideration pixel thermalization times, Monte-Carlo simulations performed with \textit{Geant4}~\cite{ALLISON2016186,ALLISON2006,AGOSTINELLI2003250} indicate the particle hit rate from these standard bunches must be reduced by at least a factor of 100 in order to avoid overwhelming signal pile-up in the PAX TES detector.

\begin{figure}[htbp!]
    \centering
    \begin{minipage}{0.5\textwidth}
        \centering
        \includegraphics[height=6cm]{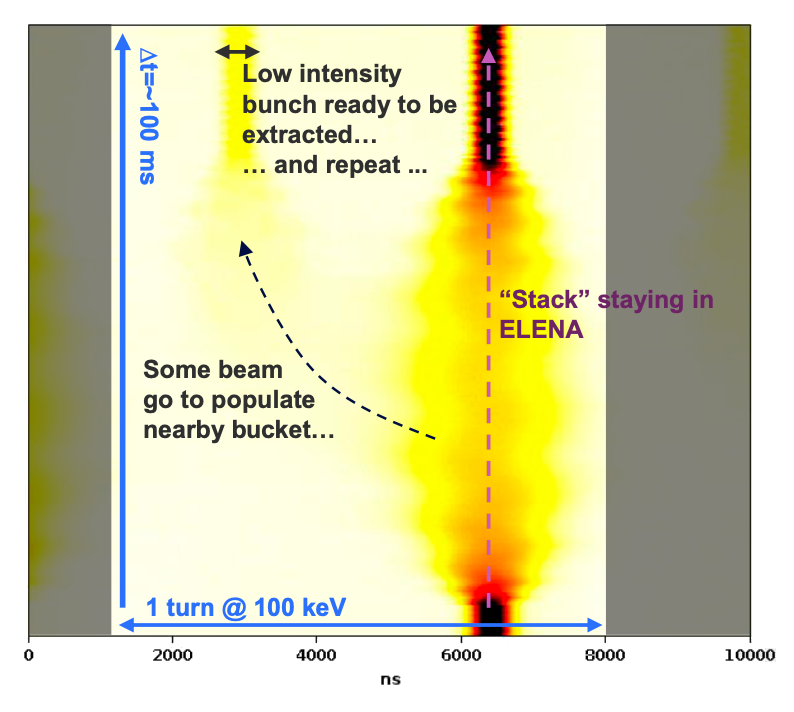}
    \end{minipage}\hfill
    \begin{minipage}{0.5\textwidth}
        \centering
        \includegraphics[height=6cm]{{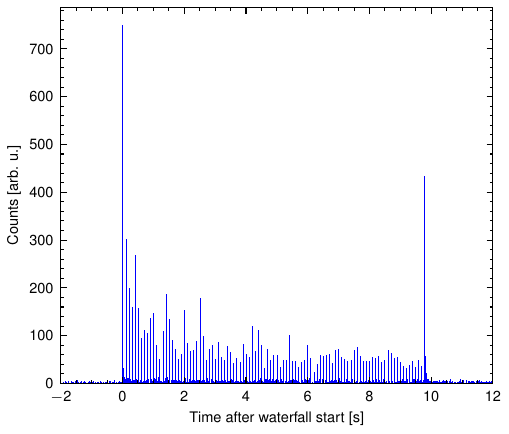}}
    \end{minipage}
    \caption{(Left) Waterfall acquisition of the longitudinal beam profile evolution in ELENA at extraction energy during proof-of-principle tests for creating a low-intensity bunch: A single bunch is partially debunched and then rebunched at the revolution frequency harmonic 2. The low-intensity tail of the original bunch is trapped in the second bucket, ready to be extracted. (Right) A scintillator spectrum of 100 small microbunches being delivered to the experimental area. The zero time is defined as the start of the spill. The large spike at \SI{10}{\second} corresponds to the remainder of the standard bunch being delivered.} 
    \label{fig:waterfall-schottky_alt}
\end{figure}

In lieu of a true slow extraction with low intensity and no bunching, a new beam extraction method is being developed for PAX by the ELENA beam physics team. 
This method involves breaking a standard ELENA bunch into smaller microbunches and delivering them one-at-a-time to the experimental zone. 
A proof-of-principle demonstration of "waterfall" extraction is shown in Figure~\ref{fig:waterfall-schottky_alt}, showing a low-intensity bunch generated in the ELENA ring spilling a few antiprotons from a high-intensity bunch, and an example of extraction every \SI{100}{\milli\second}.
Figure~\ref{fig:waterfall-schottky_alt} also shows the multi-bunch extraction as delivered to the TELMAX experimental zone at ELENA and measured with a large paddle scintillator. The 100 microbunches can be seen over the \SI{10}{\second} spill.
The waterfall extraction method allows for detailed control over microbunch intensity and delivery timing, both of which will be optimized based on the observed rates in the TES.  

\section{Summary and Perspectives}
PAX presents a novel methodological approach to BSQED tests that leverages the confluent availability of antimatter beams at CERN's ELENA ring and the now mature technology of transition edge sensing x-ray detectors. 
The high precision and detection efficiency of large area TESs will enable PAX to perform antiprotonic x-ray spectroscopy of transitions between Rydberg states in antiprotonic atoms at the $10^{-5}$--$10^{-6}$ accuracy level, accessing for the first time second-order QED effects in these systems. These represent the strongest-field bound-state atomic systems that can be created in the laboratory and will open up a new frontier for precision studies of the quantum vacuum. The PAX measurements will be complementary to ongoing work with HCIs and muonic atoms, and when highest accuracy measurements are achieved, may allow to place additional constraints on new baryonic interactions with decays to the dark sector. In the longer term, PAX will be able to contribute to other studies with antiprotonic atoms, notably by providing antiprotonic x-ray cascade data, complementary to pion annihilation signals, to help study neutron skins of exotic nuclei within the PUMA experiment \cite{Obertelli2018,Aumann2022}.

\section*{Acknowledgements}

The PAX project is supported for the period of 2023-2024 by the French ANR Young Researcher's Grant (JCJC) n° ANR-23-CE30-0002. PAX is supported during the period of 2024-2029 by the European Research Council Starting Grant n° 101115849. This project has received funding from the European Union’s Horizon Europe research and innovation programme under grant agreement n° 101057511.

We would like to thank the AD/ELENA technical staff at CERN for the availability of the new TELMAX area, which has allowed first beam tests for this project to be possible, and  J. Weber of CU Boulder for providing the Drawings of Figure~\ref{fig:tes-drawing}. 

\bibliographystyle{apsrev}

\bibliography{bibliography.bib}

\end{document}